\documentclass[12pt]{article}
 \usepackage{graphicx}
 \usepackage[cp1251]{inputenc} 

\begin{document}
 \noindent {\footnotesize\it Astronomy Letters, 2016, Vol. 42, No. 3, pp. 182--192.}
 \newcommand{\dif}{\textrm{d}}

 \noindent
 \begin{tabular}{llllllllllllllllllllllllllllllllllllllllllllll}
 & & & & & & & & & & & & & & & & & & & & & & & & & & & & & & & & & & & & & \\\hline\hline
 \end{tabular}

  \vskip 0.5cm
  \centerline{\bf The z Distribution of Hydrogen Clouds and  Masers}
  \centerline{\bf with Kinematic Distances}
  \bigskip
  \centerline{V.V. Bobylev and A.T. Bajkova}
  \bigskip
  \centerline{\small\it Pulkovo Astronomical Observatory, St. Petersburg,  Russia}
  \bigskip
  \bigskip
{\bf Abstract}—Data on HII regions, molecular clouds, and methanol
masers have been used to estimate the Sun's distance from the
symmetry plane $z_\odot$ and the vertical disk scale height $h.$
Kinematic distance estimates are available for all objects in
these samples. The Local-arm (Orion-arm) objects are shown to
affect noticeably the pattern of the $z$ distribution. The
deviations from the distribution symmetry are particularly
pronounced for the sample of masers with measured trigonometric
parallaxes, where the fraction of Local-arm masers is large. The
situation with the sample of HII regions in the solar neighborhood
is similar. We have concluded that it is better to exclude the
Local arm from consideration. Based on the model of a
self-gravitating isothermal disk, we have obtained the following
estimates from objects located in the inner region of the Galaxy
$(R\leq R_0): z_\odot= -5.7\pm0.5$~pc and $h_2=24.1\pm0.9$~pc from
the sample of 639 methanol masers, $z_\odot=-7.6\pm0.4$~pc and
$h_2=28.6\pm0.5$~pc from 878 HII regions, $z_\odot=-10.1\pm0.5$~pc
and $h_2=28.2\pm0.6$~pc from 538 giant molecular clouds.


\section*{INTRODUCTION}
The Galactic thin disk attracts the attention of many authors.
Data on young O- and B-type stars, open clusters, Cepheids,
infrared sources, molecular clouds, and other young objects are
used to study its properties. In particular, such parameters as
the Sun's distance from the symmetry plane $z_\odot$ and the
vertical disk scale height $h$ are important. Note that two terms
are used: (a) the Sun's height above the Galactic plane and (b)
the Sun's position relative to the symmetry plane. The Sun is
known to slightly rise above the Galactic plane; therefore, the
Sun's height is positive and is usually designated as $h_\odot$.
In the heliocentric reference frame, the Sun's position relative
to the symmetry plane of the Galaxy is calculated as the mean of
the $z$ coordinates of surrounding objects and, therefore, is
negative at a positive height. This distance is usually designated
as $z_\odot$.

The results of the analysis of samples with the distances
estimated from trigonometric parallaxes (Maiz-Apell\'aniz 2001),
photometrically (Reed 2000), by analyzing open star clusters
(Piskunov et al. 2006; Elias et al. 2006; Joshi 2007; Buckner and
Froebrich 2014), or from a combination of various data (Bobylev
and Bajkova 2016) are primarily of interest. At the same time, the
less accurate kinematic distances have not lost their significance
in analyzing the youngest disk objects, such as hydrogen clouds or
methanol masers. Using such distances allows large samples of
objects distributed almost over the entire Galactic disk to be
analyzed (Bronfman et al. 2000; Fish et al. 2003; Paladini et al.
2004; Pandian et al. 2009).

Previously (Bobylev and Bajkova 2016), we estimated $z_\odot$ and
$h$ using OB associations, HII regions, Wolf–Rayet stars with
known photometric distances, and a sample of classical Cepheids
with their distances estimated from the period–luminosity
relation. In addition, we used a sample of 90 masers with their
trigonometric parallaxes measured by VLBI. The errors in these
distances are very small, on average, 10\%. However, when studying
this sample, we found that the distribution of masers in a plane
perpendicular to the Galactic plane has a noticeable asymmetry
compared to the Gaussian one. In this paper, we want to ascertain
what caused this asymmetry using various samples and various
constraints. The goal of this paper is to minimize the influence
of the main contributors (the Local arm and the Galactic disk
warp) that give rise to an asymmetry in the $z$ distribution of
young Galactic-disk objects and to estimate the vertical disk
scale height $h.$ To solve this problem, we use a large
statistical material from the compilation of Hou and Han (2014),
which contains information about more than 2500 HII regions, 1300
giant molecular clouds, and 900 methanol masers with the estimates
of their kinematic distances.

\section*{METHODS}
In this paper, we use two rectangular coordinate systems: the
moving heliocentric $xyz$ and fixed Galactocentric $XYZ$ ones. In
the heliocentric $xyz$ coordinate system, the $x$ axis is directed
toward the Galactic center, the $y$ axis is in the direction of
Galactic rotation, and the $z$ axis is directed to the north
Galactic pole. When we talk about the Galactic quadrants, we mean
precisely this coordinate system. Thus, quadrants I, II, III, and
IV span the ranges of Galactic longitudes $0^\circ<l\leq 90^\circ,
90^\circ<l\leq 180^\circ, 180^\circ<l\leq 270^\circ,$ and
$270^\circ<l\leq 360^\circ,$ respectively.

In the Galactocentric $XYZ$ coordinate system, the $X$ axis is
directed to the object from the Galactic center, the $Y$ axis
coincides with the direction of Galactic rotation, and the $Z$
axis is directed toward the north Galactic pole. Here, it is
important to know the Galactocentric distance R0 of the Sun. We
use the present-day recent estimate of $R_0=8.34\pm0.16$~kpc
obtained by Reid et al. (2014) from a large sample of Galactic
masers with their trigonometric parallaxes measured by VLBI. In
this coordinate system, it is convenient, for example, to draw the
spiral pattern or to specify the inner $(R\leq R_0)$ and outer
$(R>R_0)$ regions of the Galaxy.

In the case of an exponential density distribution, the observed
histogram of the distribution of objects along the $z$ coordinate
axis is described by the expression
 \begin{equation}
  N(z)=N_1 \exp \biggl(-{|z-z_\odot|\over h_1} \biggr),
 \label{elliptic}
 \end{equation}
where $N_1$ is the normalization coefficient and $z_\odot$ is the
Sun's distance from the symmetry plane.

If the model of a self-gravitating isothermal disk is used for the
density distribution, then the observed frequency distribution of
objects along the $z$ axis is described by
 \begin{equation}
  N(z)=N_2{\hbox { sech}}^2 \biggl({z-z_\odot\over \sqrt2~h_2}\biggr).
 \label{self-grav}
 \end{equation}
When comparing the results, it should be kept in mind that
different authors use differing coefficients in the denominator
when applying model~(2): either two,
 $N(z)=N_2{\hbox { sech}}^2 [(z-z_\odot)/2h]$
 (Maiz-Apell\'aniz 2001; Buckner and Froebrich 2014), or one,
 $N(z)=N_2{\hbox { sech}}^2 [(z-z_\odot)/h]$ (Marshall et al. 2006).

Finally, the observed frequency distribution of objects along the
$z$ axis for the Gaussian model is described by the formula
 \begin{equation}
  N(z)=N_3\exp\biggl[-{1\over 2}\biggl({z-z_\odot\over h_3}\biggr)^2\biggr].
 \label{Gauss}
 \end{equation}
As was shown, for example, by Maiz-Apell\'aniz (2001), $h$ is
determined with the smallest errors based on Eq.~(2); therefore,
this model is most attractive among the three described ones.

When analyzing the distributions of OB stars and open star
clusters, many authors (Stothers and Frogel 1974; Reed 2000; Elias
et al. 2006; Joshi 2007) made great efforts to rid the sample of
the influence of the Gould Belt. An overview of the Gould-Belt
properties can be found, for example, in Bobylev (2014). However,
the fraction of Gould-Belt objects is quite small in the samples
of HII regions or methanol masers. For example, only 12 masers in
the sample of 130 masers with measured trigonometric parallaxes
belong to the $r<0.5$~kpc neighborhood, where the Gould Belt is
located. However, the Gould Belt entirely belongs to a larger
structure, the Local arm (Orion arm).

Previously (Bobylev and Bajkova 2016), we showed the $z$
distribution of 130 masers with measured trigonometric parallaxes
to be asymmetric. We suggested that this is because the sample is
dominated by the masers observed mainly from the Earth's northern
hemisphere. However, there can also be other sources of this
peculiarity, which should be studied using a large statistical
material. In particular, it is interesting to check the influence
of the Local arm. As we showed previously (Bobylev and Bajkova
2014a), the symmetry plane of the Local arm has an inclination of
about $6^\circ$ to the Galactic plane. Since about 40 of the 130
masers belong to the Local arm, the effect here can be
significant.

 \begin{figure} {\begin{center}
 \includegraphics[width=120mm]{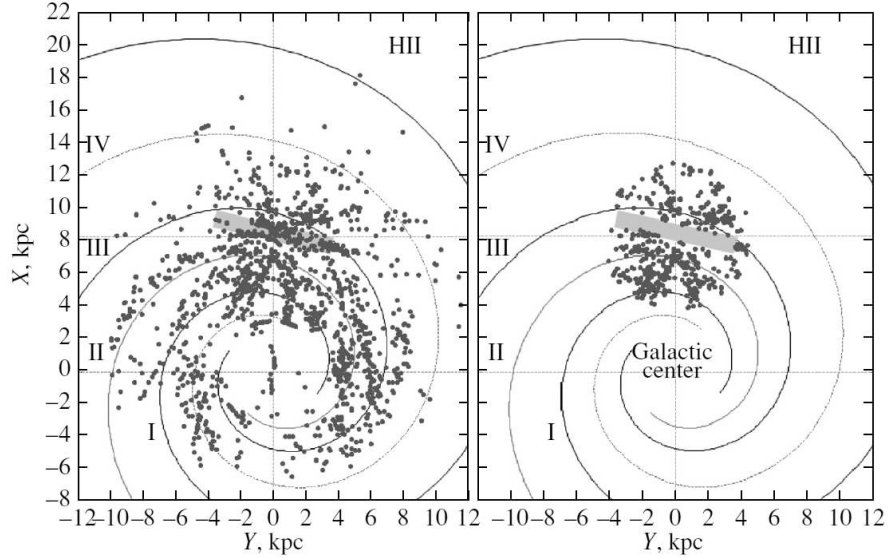}
 \caption{Distributions of the sample of 1771 HII regions (left) and the sample
of 653 HII regions in the solar neighborhood, $r<4.5$~kpc, without
Local-arm objects (right) in the Galactic $X$Y plane. The Sun's
coordinates are $(X,Y)=(8.3,0)$~kpc. The four-armed spiral pattern
with a pitch angle of $-13^\circ$ is plotted; the spiral arms are
numbered by Roman numerals; the gray rectangle indicates the
Local-arm model. }
 \label{fig1-XY} \end{center} } \end{figure}
 \begin{figure} {\begin{center}
 \includegraphics[width=120mm]{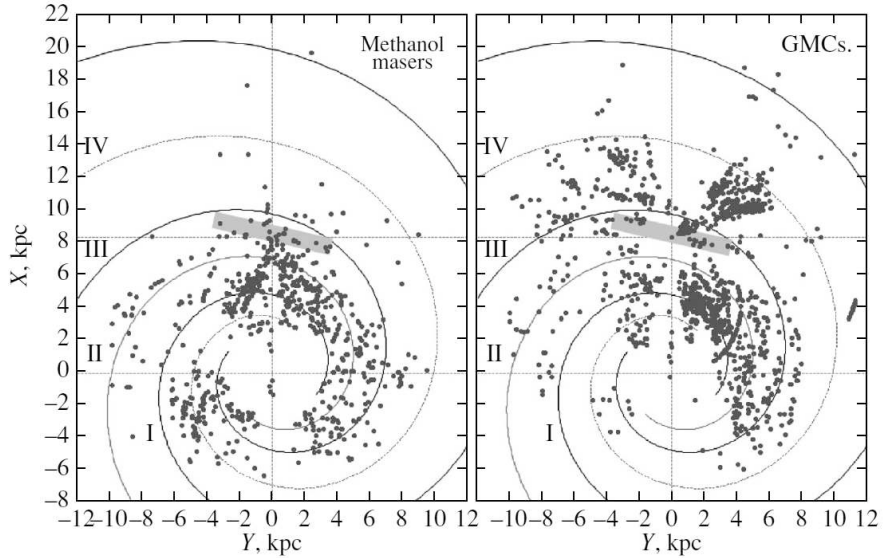}
 \caption{Distributions of the sample of 716 methanol masers (left) and the sample of 1281
giant molecular clouds (right) in the Galactic $XY$ plane. The
Sun's coordinates are $(X,Y)=(8.3,0)$~kpc. The four-armed spiral
pattern with a pitch angle of $-13^\circ$ is plotted; the spiral
arms are numbered by Roman numerals; the gray rectangle indicates
the Local-arm model. }
 \label{fig2-XY} \end{center} } \end{figure}

\section*{DATA}
We use the data on $\sim$1800 HII regions, 1300 giant molecular
clouds (GMCs), and $\sim$700 methanol masers from Hou and Han
(2014), for which these authors estimated the distances
kinematically. The distribution of all HII regions with available
kinematic distance estimates in the Galactic $XY$ plane is
presented in Fig.~1, while the distribution of methanol masers and
GMCs is presented in Fig.~2.

Figures 1 and 2 show a fragment of the global four-armed Galactic
spiral pattern with a constant pitch angle of $-13^\circ$. The
pattern was constructed in our previous paper (Bobylev and Bajkova
2014b), where we analyzed the distribution of Galactic masers with
measured trigonometric parallaxes, with the pattern parameters
having been recalculated for the Galactocentric distance of the
Sun adopted in this paper, $R_0=8.3$~kpc. As can be seen from
these figures, the distribution of objects agrees satisfactorily
with the plotted spiral pattern. In all samples, we see that the
objects fall nicely on the segment of the Carina--Sagittarius arm
(arm II in the figures) in forth quadrant. This is seen best in
the distribution of HII regions. Note that Hou and Han (2014) also
reached the conclusion that the Galaxy most likely has a
four-armed spiral pattern with a pitch angle of $-13^\circ$. At
the same time, they concluded that the model of a four-armed
pattern with a variable pitch angle is even more suitable.

There are photometric distance estimates obtained by different
authors for a relatively small number ($\sim$200) of HII regions
in the compilation of Hou and Han (2014). These are objects fairly
close to the Sun. Their distribution resembles the distribution of
HII regions within 4.5~kpc of the Sun (the right panel in Fig.~1).
Hou and Han (2014) added about a hundred more masers with measured
trigonometric parallaxes to their sample of HII regions. These are
not only methanol (CH$_3$OH) masers but also H$_2$O and SiO masers
observed in the radio band at various frequencies using VLBI. The
accuracy of such measurements is, on average, about 10\%.

In this paper, we use a more complete sample of masers with
measured trigonometric parallaxes that contains 130 sources. We
compiled this sample based on a number of publications. The main
work is the review by Reid et al. (2014), where the data on 103
masers are described. Subsequently, the publications of these
authors with improvements and supplements devoted to the analysis
of masers located in individual spiral arms of the Galaxy
appeared. These include Sanna et al. (2014), Sato et al. (2014),
Wu et al. (2014), Choi et al. (2014), and Hachisuka et al. (2015).
Note the paper of Xu et al. (2013) devoted entirely to the masers
in the Local arm. We used this sample of 130 masers previously
(Bobylev and Bajkova 2016).

\section*{RESULTS}
\subsection*{Influence of the Local Arm}
As can be seen from Figs. 1 and 2, the Local arm is represented by
a considerable number of objects only in the distribution of HII
regions. In contrast, the number of Local-arm objects in the
distribution of methanol masers is small, while there are
virtually no Local-arm objects among the GMCs.

According to Bobylev and Bajkova (2014a), we took a
$6.2\times1.1$~kpc rectangle oriented at an angle of $-13^\circ$
to the y axis and displaced from the Sun by 0.3 kpc toward the
Galactic anticenter as the simplest Local-arm model. In accordance
with this model, we cut out the HII regions on the right panel of
Fig.~1, as marked by the gray rectangle located between the
Perseus (arm III) and Carina--Sagittarius (arm II) arms.

Figure 3 presents the histograms of the $z$ distribution for HII
regions with kinematic distances and the sample of masers with
measured trigonometric distances. The light shading on the left
panel of the figure indicates the distribution for the sample of
653 HII regions from the $r<4.5$~kpc solar neighborhood that
contains no Local-arm objects, while the darker shading indicates
the distribution of 134 HII regions belonging to the Local arm.
The light shading on the right panel of the figure indicates the
distribution for the sample of 88 masers that contains no
Local-arm objects, while the thick line marks the distribution for
the sample of 42 Local-arm masers.

 \begin{figure} {\begin{center}
 \includegraphics[width=160mm]{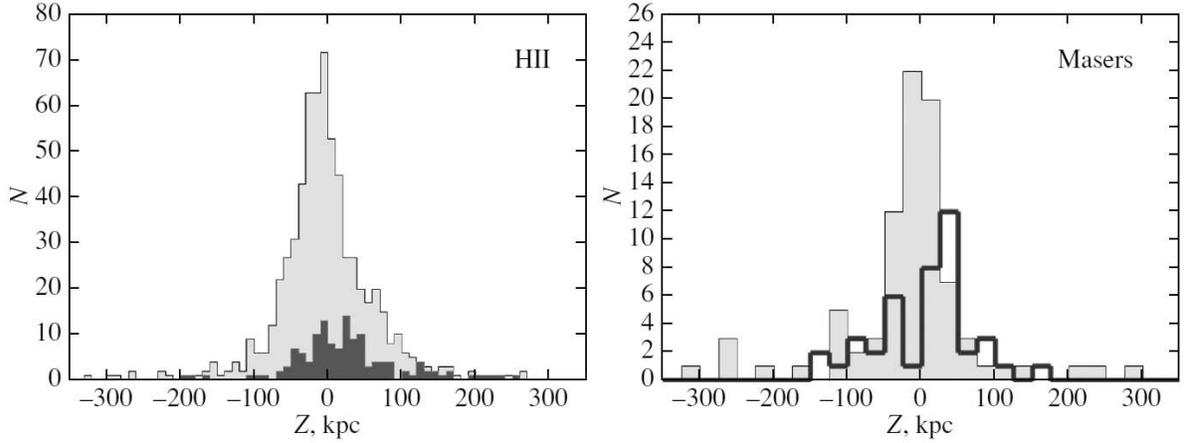}
 \caption{Histograms of the $z$ distributions of HII regions and the sample of
masers with measured trigonometric parallaxes. The dark shading on
the left panel and the thick dark line on the right panel
highlight the Local-arm objects. }
 \label{Local-Arm} \end{center} } \end{figure}

Based on 653 HII regions from the $r<4.5$~kpc solar neighborhood
that contains no Local-arm objects, we found
 \begin{equation}
 z_\odot= -5.1\pm2.6~\hbox{pc},
 \label{rez-1}
 \end{equation}
where $z_\odot$ was calculated as a simple mean, and its error was
found as the error of the mean. Based on 134 HII regions belonging
to the Local arm, we obtain
 \begin{equation}
 z_\odot= +26.7\pm6.5~\hbox{pc}.
 \label{rez-2}
 \end{equation}
We can see a significant difference. We found
 \begin{equation}
 z_\odot= -14.3\pm11.6~\hbox{pc},
 \label{rez-3}
 \end{equation}
based on the sample of 88 masers (without any constraints on the
distance from the Sun) that contains no Local-arm objects and
 \begin{equation}
 z_\odot= +4.7\pm9.9~\hbox{pc},
 \label{rez-4}
 \end{equation}
based on 42 Local-arm masers, and here the difference is also
tangible. This difference is particularly clearly seen on the
corresponding panel of Fig.~3. Finally, if the masers from the
0.5-kpc neighborhood (Gould-Belt masers) are rejected, then from
the remaining 29 Local-arm masers we find
 \begin{equation}
 z_\odot= +23.6\pm12.6~\hbox{pc}.
 \label{rez-5}
 \end{equation}
i.e., the entire Local arm gives a shift in the positive direction
when calculating the mean $z_\odot$. Given the height of the Sun
above the Galactic plane, we can conclude that the entire Local
arm rises above the Galactic plane by about 25--35 pc.

Note that previously (Bobylev and Bajkova 2015) we revealed
appreciable ($\sim$10 km s$^{-1}$) velocities in a direction
perpendicular to the Galactic plane (the velocities $W$ directed
along the $Z$ axis). The Galactic spiral density wave is probably
responsible for such periodic perturbations. The perturbations
turned out to be particularly pronounced in the regions of the
Gould Belt and the Orion and Perseus arms. On the whole, we can
conclude that it is better not to use the Local-arm objects in the
sample when analyzing the $z$ distributions of stars. Below, we
apply this rule to all our samples, in particular, to all samples
in the table.

\subsection*{The $r<4.5$~kpc Neighborhood}
The table presents the results of determining the Sun’s height
$z_\odot$ and the vertical disk scale height $h_i, i=1,2,3$ based
on models (1), (2), and (3) using various samples of stars. These
parameters and their errors were found by fitting the models to
the histograms and Monte Carlo simulations. For this purpose, we
constructed the histograms with a step in $z$ of 10 pc. We assumed
the errors in the distances determined kinematically to be 30\%
for all our samples. The upper part of the table presents the
results obtained from the samples of objects that are located in
the solar neighborhood with a radius of 4.5 kpc and contain no
Local-arm objects.

Figure 4 presents the histograms of the $z$ distributions for 240
masers, 653 HII regions, and 364 GMCs, where the curves of models
(1) and (2) constructed with the parameters specified in the table
are plotted. As can be seen from the figure, there is good
agreement between the two model curves for all three samples. The
distributions of masers and HII regions are quite symmetric and
have no appreciable deviations from the Gaussian. In this regard,
the distribution of GMCs appears farther from the Gaussian one,
which may be because their longitude distribution is clumpy.

Figure 5 presents the longitude distribution of masers, HII
regions, and GMCs. These are the samples used to find the
parameters specified in the table and to construct Fig.~4. As can
be seen from the figure, almost all of the masers lie at
$|l|<30^\circ$, because they are located in the inner region of
the Galaxy. In the sample of masers, there is virtually no
influence of the disk warp. In the two other samples, the
influence of the disk warp manifests itself as a sine wave
especially pronounced in the distribution of HII regions. This can
be associated with large random errors in the distances. We used
the constraint $r=4.5$~kpc at which, as we expected, the influence
of the disk warp should not be noticeable. However, because of the
errors in the distance estimates (probably more than 30\%), more
distant objects that are significantly affected by the Galactic
disk warp penetrated the sample.

The right panels of Fig.~5 present the Local-arm objects, while
the probable members of the Gould Belt (selected from the solar
neighborhood with a radius $r<0.6$~kpc) are highlighted for the
HII regions. As can be seen from the figure, HII regions are the
most numerous members of the Local arm. 178 HII regions belong to
the Local arm; 45 of them are located at $r<0.6$~kpc. The fact
that there are quite a few HII regions with large positive $z$ in
the range of longitudes $180^\circ<l<240^\circ$ is unexpected. At
least for the Gould-Belt objects in this range, one might expect
negative $z$ (as, for example, for the members of the Orion OB
association). By contrast, the distribution of nearby GMCs that
form two clumps near $l\approx90^\circ$ and $l\approx240^\circ$
agrees well with the expected picture.

 \begin{table}[t]                                     
 \caption[]{\small
 The Sun's height $z_\odot$ and the vertical disk scale height $h_{i, ~i=1,2,3}$
 obtained using models (1), (2), and (3) from three samples with kinematic distances}
  \begin{center}  \label{t:03}   
  \begin{tabular}{|c|c|c|c|l|c|}\hline
 $z_\odot,$~pc & $h_1,$~pc & $h_2,$~pc & $h_3,$~pc & Sample   \\\hline

 $~-6.4\pm0.9$ & $21.4\pm1.1$ & $20.7\pm1.8$ & $21.1\pm1.4$ & 240 masers,  ~~~~~~$r<4.5$~kpc \\
 $~-5.1\pm0.8$ & $41.5\pm0.8$ & $36.4\pm0.7$ & $42.0\pm0.9$ & 653 HII regions, ~$r<4.5$~kpc \\
 $-12.2\pm1.1$ & $47.9\pm1.5$ & $42.6\pm1.3$ & $49.5\pm1.6$ & 364 GMCs,  ~~~~~~~$r<4.5$~kpc \\
 \hline
 $~-5.7\pm0.5$ & $26.5\pm0.7$ & $24.1\pm0.9$ & $25.9\pm0.6$ & 639 masers,  ~~~~~~~$R\leq R_0$ \\
 $~-7.6\pm0.4$ & $32.8\pm0.6$ & $28.6\pm0.5$ & $33.0\pm0.6$ & 878 HII regions, ~$R\leq R_0$ \\
 $-10.1\pm0.5$ & $33.8\pm0.7$ & $28.2\pm0.6$ & $32.5\pm0.7$ & 538 GMCs,  ~~~~~~~$R\leq R_0$ \\
 \hline
 $~+0.9\pm1.8$ & $ 93.7\pm2.6$ & $83.9\pm2.3$ & $ 98.7\pm2.9$ & 380 HII regions, ~$R> R_0$ \\
 $~+2.4\pm1.6$ & $110.5\pm2.9$ & $95.0\pm2.3$ & $110.1\pm2.8$ & 474 GMCs,  ~~~~~~~$R> R_0$ \\
 \hline
 \end{tabular}\end{center}
 \end{table}
 \begin{figure} {\begin{center}
 \includegraphics[width=80mm]{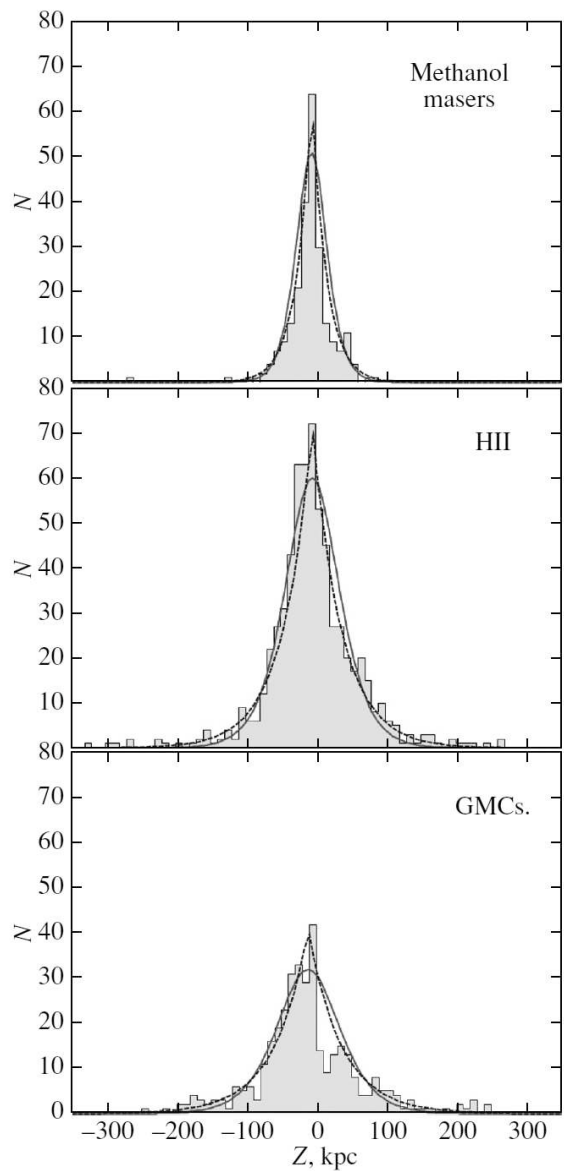}
 \caption{Histograms of the $z$ distributions of 240 methanol masers (top),
 653 HII regions (middle), and 364 giant molecular
clouds (bottom). The dashed and solid lines on all panels
represent models (1) and (2), respectively. All samples are
located in the $r<4.5$~kpc neighborhood and contain no Local-arm
objects.}
 \label{f-hist-all} \end{center} } \end{figure}
 \begin{figure} {\begin{center}
 \includegraphics[width=160mm]{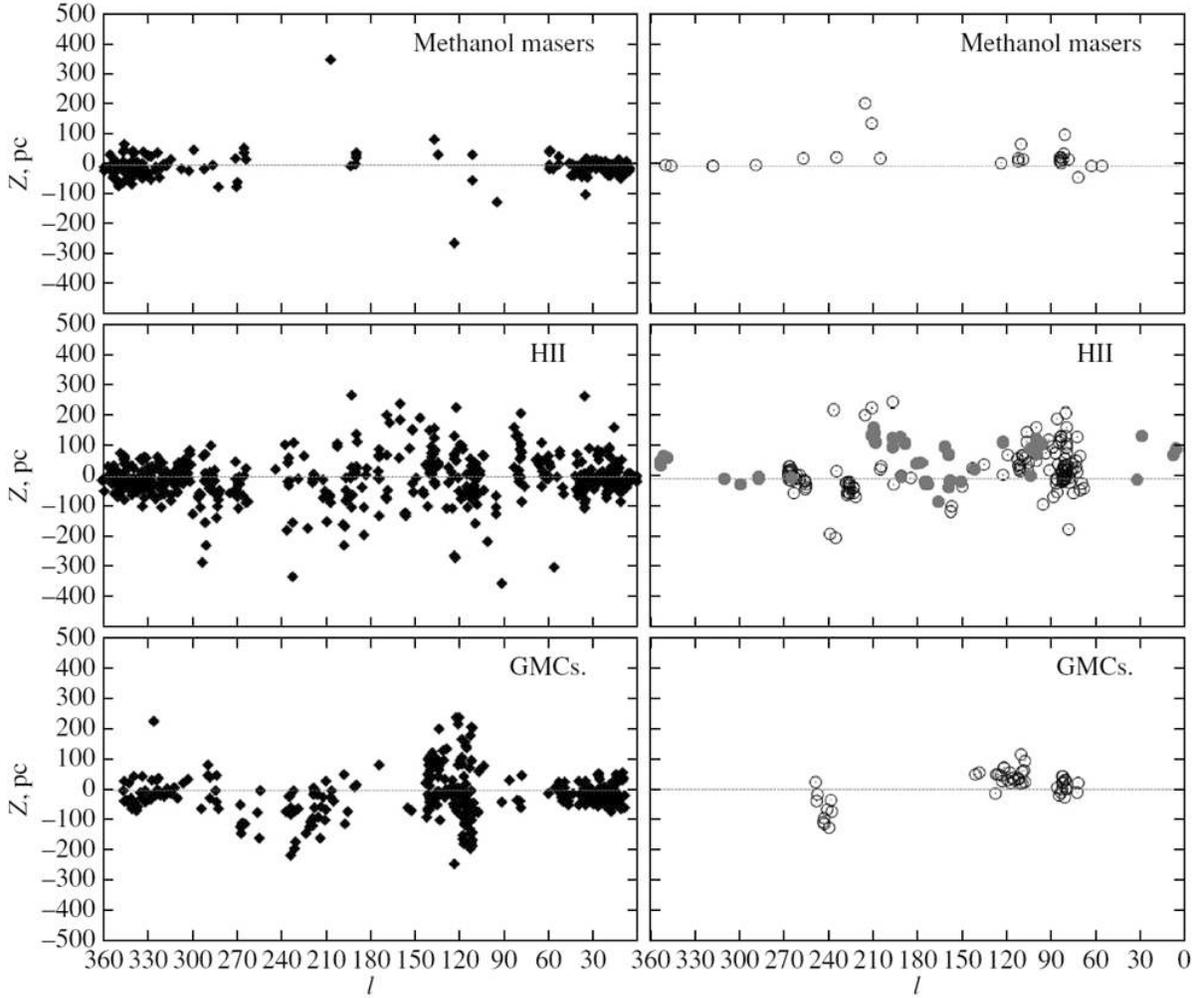}
 \caption{Longitude distributions of 240methanolmasers (top left),
 653 HII regions (middle left), and 364 giant molecular clouds
(bottom left). These samples are located in the $r<4.5$~kpc solar
neighborhood and contain no Local-Arm objects. The right panels
present the Local-arm objects (open circles), while the probable
Gould-Belt objects are highlighted (filled circles) for the HII
regions. }
 \label{LB-all} \end{center} } \end{figure}
 \begin{figure} {\begin{center}
 \includegraphics[width=160mm]{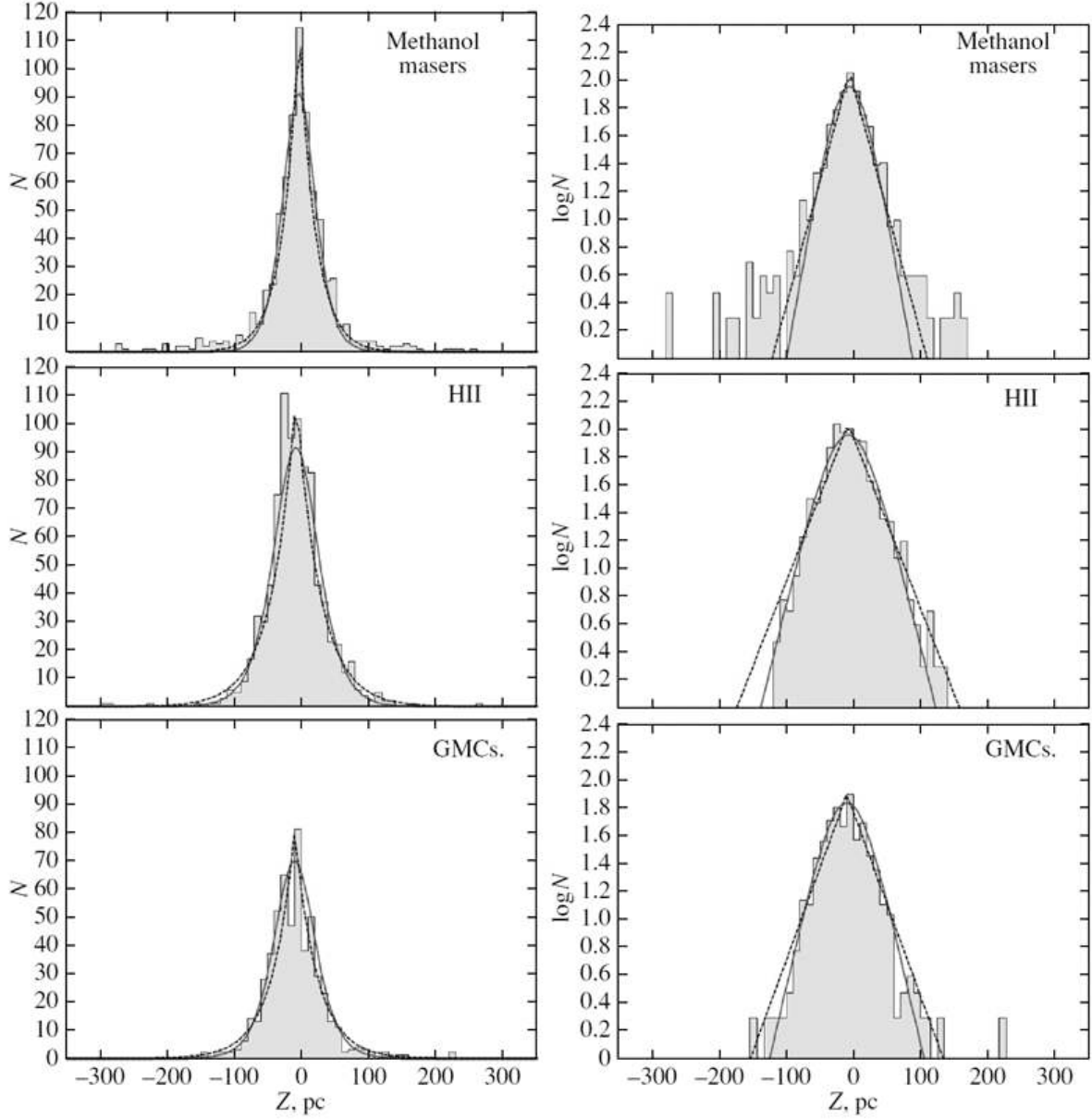}
 \caption{(a) Histograms of the $z$ distributions of 639 methanol masers,
 878 HII regions, and 538 GMCs from the inner region
of the Galaxy. (b) The same histograms on a logarithmic scale. The
dashed and solid lines on all panels represent models (1) and (2),
respectively.}
 \label{hist-all-R0} \end{center} } \end{figure}

\subsection*{The Inner and Outer Regions of the Galaxy}
When analyzing the distances estimated kinematically from the
Galactic rotation curve using the line-of-sight velocities, the
separation into the inner and outer regions is justified by the
form of errors. As Sofue (2011) showed, the distribution of errors
in the distances when estimated via the line-of-sight velocities
has a complex shape like a butterfly: its ``body'' located at the
Galactic center is elongated in the Galactic center--Sun direction
(the $X$ axis), while its wings are elongated along the $Y$ axis.
The inner regions of the Galaxy have the smallest errors, except
for the narrow zone around the tangential circumference.

Note that the $XY$ distribution of almost the entire sample of
methanol masers (Fig.~2) is close to the region of small errors in
the kinematic distances. Almost the entire sample, 639 methanol
masers, lies in the inner region of the galaxy, $R\leq R_0.$ In
contrast, there are few objects, about 130 masers, in the outer
region of the Galaxy, $R>R_0,$ and, therefore, we do not use them.

Figure 6 presents the histograms of the $z$ distributions of 639
methanol masers, 878 HII regions, and 538 GMCs from the inner
region of the Galaxy, $R\leq R_0.$ The condition that allows the
Local-arm region to be excluded was applied to all our samples.
The criterion $r<10$~kpc was additionally applied for the HII
regions and GMCs, which allowed the very distant objects to be
rejected. All three histograms in this figure are also presented
on a logarithmic scale. The Sun's height and the vertical disk
scale height obtained from all these samples are given in the
middle and lower parts of the table.

Obviously, the results presented in the two lower rows of the
table are far from the characteristics of the young disk. The
positive $z_\odot$ and the huge $h$ point to a very strong
influence of the disk warp.

\section*{DISCUSSION}
Comparison of the parameters presented in the table and the
constructed z distributions in Figs.~4 and 6 shows an advantage of
the samples located in the inner region of the Galaxy. For
example, the distribution of GMCs in Fig.~6 looks considerably
more symmetric than that in Fig.~4; the distributions of HII
regions and methanol masers became closer to the Gaussian. The
closeness of the distributions to the theoretical ones specified
by Eqs.~(1), (2), and (3) was estimated each time using the
$\chi^2$ test from the formula $\chi^2=\Sigma_i
(\theta_i-\Phi_i)^2/\Phi_i$, where $\theta_i$ are the sampled
values of the histogram normalized in such a way that $\Sigma_i
\theta_i=1$, $\Phi_i$ is the theoretical probability distribution
that also obeys the normalization condition $\Sigma \Phi_i=1$. For
example, the $\chi^2$ estimates for the approximations by
distributions~(1), (2), and (3) were 0.284, 0.837, and 0.999,
respectively, for the sample of 364 GMCs (Fig.~4) and 0.197,
0.121, and 0.258 for the sample of 538 GMCs (Fig.~6). Comparison
of these estimates suggests a more accurate approximation of the
data by the theoretical distributions for the sample of 538 GMCs
(Fig.~6) than that for the sample of 364 GMCs (Fig.~4). This is
particularly clearly seen for the approximation by
distribution~(2).

The HII regions, giant molecular clouds, and methanol masers are
intimately connected between themselves. All these objects are
very young (approximately of the same age) and enter into star
forming regions. One might expect the vertical disk scale heights
found from these samples to be close. However, we can see such a
situation only in the middle part of the table.

Note that many authors determined the Sun's height above the
Galactic plane and the vertical disk scale height from various
samples of HII regions and from molecular clouds. Masers were used
much more rarely. Reviews of the $z_\odot$ and $h$ determinations
from various data can be found in Humphreys and Larsen (1995),
Reed (2006), or Bobylev and Bajkova (2016). Therefore, the results
that we obtained from a huge sample of methanol masers are of
considerable interest. After all, methanol masers are associated
with massive protostars in regions of active star formation, and,
consequently, they reflect the distribution of just born massive
stars.

Bronfman et al. (2000) studied the $z$ distribution of 748
star-forming regions and H$_2$ clouds with available kinematic
distances. These authors found $z_\odot\approx5$~pc and
$h\approx40$~pc (here, $h$ is the half-width at half maximum of
the Gaussian distribution) from HII regions and
$z_\odot\approx-6$~pc and $h\approx60$~pc from molecular hydrogen
clouds. They used the inner region of the Galaxy, $R/R_0=0.2-1.0.$
Nevertheless, they found significant differences in the $z_\odot$
determinations from objects in quadrants I and II
$(0^\circ<l<180^\circ)$ compared to objects in quadrants III and
IV $(180^\circ<l<360^\circ).$ For example, they found
$z_\odot=-12$~pc and $z_\odot=+3$~pc from northern- and
southern-sky molecular clouds, respectively.

Based on a sample of 20 ultracompact HII regions, by first
estimating their kinematic distances, Fish et al. (2003) found the
vertical scale height to be $34.7\pm1.4$~pc (for the derived
$z_\odot=-7.3\pm1.4$~pc). Note that the above scale height is the
half-width at half maximum of the Gaussian.

Paladini et al. (2004) analyzed the $z$ distribution of 550 HII
regions. The distances to these regions were estimated in part
kinematically from the Galactic rotation curve and in part based
on the luminosity--diameter correlation. These authors found
$z_\odot=-11.3$~pc and $h_3=52$~pc for the inner region of the
Galaxy, $R<R_0.$

Pandian et al. (2009) determined the kinematic distances for 86
methanol masers and found the vertical scale heights of the masers
to be $h_1=20$~pc $(z_\odot=-12.8$~pc) and $h_3=30$~pc
$(z_\odot=-13.1$~pc). In their opinion, these vertical scale
heights are considerably smaller than those for the Galactic thin
disk as a whole. Moreover, these authors also analyzed the $z$
distribution for a sample of dark infrared clouds and found
complete agreement of their $h$ value with that found from masers.

Note the paper of Wienen et al. (2015), where the authors
estimated the kinematic distances to 689 complexes of the ATLASGAL
program (APEX Telescope Large Area Survey; the observations were
performed at a wavelength of 870 mm) located in the first and
fourth Galactic quadrants (almost all of them are located in the
inner region of the Galaxy). The sources of this program are
associated with massive cold clouds of dust located in regions of
active star formation. It is important that a special technique
that allows one to choose between the two alternative solutions
(near or far distance) arising in this problem based on the
character of spectral lines was applied in calculating the
kinematic distances. A total of more than 1000 individual sources
combined into complexes were used. The line-of-sight velocities
were calculated as the mean from several molecular lines. Such
measures allowed a high accuracy of the results to be achieved.
These authors obtained the following estimates:
$z_\odot=-7\pm1$~pc and $h_1=28\pm2$~pc.

We see that the spread in $h$ is from 20 and 50 pc, and there are
grounds for this spread: everything depends strongly on how the
sources are distributed in the Galaxy. However, the minimum values
of hi specified in the middle part of the table probably
correspond best to the real picture.

The derived values of $z_\odot$ lying in the range from $-6$ to
$-10$~pc are in good agreement with the results of other authors
obtained from similar objects. At the same time, these values
differ noticeably from the most probable estimates computed by
using older objects (open star clusters, Cepheids, etc.). For
example, previously (Bobylev and Bajkova 2016) we calculated the
mean $z_\odot=-16\pm2$~pc based on the individual $z_\odot$
determinations from various samples.

We can conclude that our values of $h$ obtained from the samples
of HII regions, GMCs, and methanol masers (at $R\leq R_0$) reflect
the character of the vertical scale height for the youngest stars
in the Galaxy.

\section*{CONCLUSIONS}
We obtained new estimates of the Sun’s distance from the symmetry
plane $z_\odot$ and the vertical disk scale height h using data on
HII regions, giant molecular clouds, and methanol masers. For all
these objects, there are distance estimates obtained by Hou and
Han (2014) kinematically.

First, we showed that the entire structure of the Local arm
affects noticeably the pattern of the distribution along an axis
perpendicular to the Galactic plane. The entire Local arm rises
above the Galactic plane by about 25--35~pc. The deviations from
the distribution symmetry are particularly pronounced for the
sample of masers with measured trigonometric parallaxes, where the
fraction of Local-Arm masers is large. The situation with the
sample of HII regions in the solar neighborhood is similar.

As a result, we concluded that it is better to exclude the
Local-arm objects from consideration. Moreover, we found that the
sough-for parameters were determined with the smallest errors from
the objects located in the inner region of the Galaxy $(R\leq
R_0).$

We applied three models of the density distribution to all three
samples: an exponential distribution, a self-gravitating
isothermal disk, and a Gaussian density distribution. All three
models yield close values of $h;$ therefore, we give only those
found on the basis of the model of a self-gravitating isothermal
disk. The following estimates were obtained from the objects
located in the inner region of the Galaxy $(R\leq R_0): z_\odot=
-5.7\pm0.5$~pc and $h_2=24.1\pm0.9$~pc from the sample of 639
methanol masers, $z_\odot=-7.6\pm0.4$~pc and $h_2=28.6\pm0.5$~pc
from 878 HII regions, and $z_\odot=-10.1\pm0.5$~pc and
$h_2=28.2\pm0.6$~pc from 538 giant molecular clouds.

\subsection*{ACKNOWLEDGMENTS}
We are grateful to the referees for their helpful
remarks that contributed to an improvement of this paper. This
work was supported by the ``Transitional and Explosive Processes
in Astrophysics'' Program P--41 of the Presidium of Russian
Academy of Sciences.

 \bigskip{REFERENCES}\medskip
 {\small

 1. V.V. Bobylev and A.T. Bajkova, Astron. Lett. 40, 783 (2014a).

 2. V.V. Bobylev and A.T. Bajkova, Mon. Not. R. Astron. Soc. 437, 1549 (2014b).

 3. V.V. Bobylev, Astrophysics 57, 583 (2014).

 4. V.V. Bobylev and A.T. Bajkova, Mon. Not. R. Astron. Soc. 447, L50 (2015).

 5. V.V. Bobylev et al., Astron. Lett. 42, 1 (2016).

 6. L. Bronfman, S. Casassus, J.May, and L.-\AA. Nyman, Astron. Astrophys. 358, 521 (2000).

 7. A.S.M. Buckner and D. Froebrich, Mon. Not. R. Astron. Soc. 444, 290 (2014).

 8. Y.K. Choi, K. Hachisuka, M.J. Reid, Y.Xu, A. Brunthaler, K.M. Menten, and T.M. Dame,
    Astrophys. J. 790, 99 (2014).

 9. P.S. Conti and W.D. Vacca, Astron. J. 100, 431 (1990).

 10. F. Elias, J. Cabrero-Ca\~{n}o, and E.J. Alfaro, Astron. J. 131, 2700 (2006).

 11. V.L. Fish, M.J. Reid, and D.J. Wilner, Astrophys. J. 587, 701 (2003).

 12. K. Hachisuka, Y.K. Choi, M.J. Reid, A. Brunthaler, K.M. Menten, A. Sanna, and
     T.M.Dame, Astrophys. J. 800, 2 (2015).

 13. L.G. Hou and J.L. Han, Astron. Astrophys. 569, 125 (2014).

 14. R.M. Humphreys and J.A. Larsen, Astron. J. 110, 2183 (1995).

 15. Y.C. Joshi, Mon. Not. R. Astron. Soc. 378, 768 (2007).

 16. J. Maiz-Apell\'aniz, Astron. J. 121, 2737 (2001).

 17. D.J. Marshall, A.C. Robin, C. Reyl\'e, M. Schultheis, and S. Picaud,
     Astron. Astrophys. 453, 635 (2006).

 18. R. Paladini, R.D. Davies, and G. DeZotti, Mon. Not. R. Astron. Soc. 347, 237 (2004).

 19. J.D. Pandian, K.M. Menten, and P.F. Goldsmith, Astrophys. J. 706, 1609 (2009).

 20. A.E. Piskunov, N.V. Kharchenko, S. R\"oser, E. Schilbach, and R.-D. Scholz,
     Astron. Astrophys. 445, 545 (2006).

 21. B.C. Reed, Astron. J. 120, 314 (2000).

 22. B.C. Reed, J.R. Astron. Soc. Canada 100, 146 (2006).

 23. M.J. Reid, K.M. Menten, A. Brunthaler, X.W. Zheng, T.M. Dame, Y. Xu, Y.Wu,
     B. Zhang, A. Sanna, M. Sato, et al., Astrophys. J. 783, 130 (2014).

 24. A. Sanna, M.J. Reid, K.M. Menten, T.M. Dame, B. Zhang, M. Sato, A. Brunthaler,
     L. Moscadelli, and K. Immer, Astrophys. J. 781, 108 (2014).

 25. M. Sato, Y.W. Wu, K. Immer, B. Zhang, A. Sanna, M.J. Reid, T.M. Dame,
     A. Brunthaler, and K.M. Menten, Astrophys. J. 793, 72 (2014).

 26. Y. Sofue, Publ. Astron. Soc. Jpn. 63, 813 (2011).

 27. L. Spitzer, Astrophys. J. 95, 329 (1942).

 28. R. Stothers and J.A. Frogel, Astron. J. 79, 456 (1974).

 29. M. Wienen, F. Wyrowski, K.M. Menten, J.S. Urquhart, T. Csengeri, C.M. Walmsley,
     S. Bontemps, D. Russeil, at al., Astron. Astrophys. 579, 91 (2015).

 30. Y.W. Wu, M. Sato, M.J. Reid, L. Moscadelli, B. Zhang, Y. Xu, A. Brunthaler,
     K.M. Menten, T.M. Dame, and X.W. Zheng, Astron. Astrophys. 566, 17 (2014).

 31. Y. Xu, J.J. Li, M.J. Reid, X.W. Zheng, A. Brunthaler, L. Moscadelli, T.M. Dame,
     B. Zhang, Astrophys. J. 769, 15 (2013).

 }
\end{document}